\documentclass[12pt,american,fleqn]{article}
\usepackage[latin1]{inputenc}
\usepackage{a4}
\usepackage{babel}
\usepackage{graphics}

\makeatletter

\newcommand{\LyX}{L\kern-.1667em\lower.25em\hbox{Y}\kern-.125emX\spacefactor1000}

\newcommand{\boldsymbol}[1]
{\mbox{\boldmath $#1$}}

\newcommand{\lyxaddress}[1]{
  \par {\raggedright #1 
  \vspace{1.4em}
  \noindent\par}
}

\makeatother

\begin{document}

\title{On the metal--insulator transition in the two--chain model of
correlated fermions}

\author{H. J. Schulz}

\maketitle

\lyxaddress{\centering Laboratoire de Physique des Solides, Université
Paris--Sud, 91405 Orsay, France}

\begin{abstract}
The doping--induced metal--insulator transition in two--chain systems of
correlated fermions is studied using a solvable limit of the \( t-J \) model
and the fact that various strong- and weak--coupling limits of the
two--chain model are in the same phase, i.e. have the same low--energy
properties. It is shown that the Luttinger--liquid parameter \( K_{\rho } \)
takes the universal value unity as the insulating state (half--filling) is
approached, implying dominant d--type superconducting fluctuations,
independently of the interaction strength. The crossover to insulating
behavior of correlations as the transition is approached is discussed.
\end{abstract}
\newpage

Models of two parallel chains of correlated fermions are of interest for the
understanding of the physical properties of a number of systems: (i)
two--leg spin ladder systems of the type \(
\mathrm{Sr}_{\mathrm{x}}\mathrm{Ca}_{14-x}\mathrm{Cu}_{24}\mathrm{O}_{41} \)
become conducting and even superconducting under pressure,\cite{uehara96}
even though they are insulating under ambient
conditions;\cite{dagotto_ladders} (ii) the electronic structure of certain
types of carbon nanotubes\cite{ebbesen96} is described by very similar
models;\cite{kane96} (iii) two--channel quantum wires may also show
interesting interaction effects. The two--chain case, being much easier to
treat by controlled analytical and numerical methods than genuinely two-- or
three-dimensional models, is also helpful in understanding long--standing
questions about the existence of superconductivity in models of correlated
fermions.

Theoretical work on the two--chain model has either considered situations
well away from
half--filling,\cite{fabrizio_2chain,schulz_2chain,balents_2chain} or
concerned the half--filled (one electron per site) case where umklapp
scattering leads to a Mott insulator state with a spin
gap.\cite{schulz_spins,strong_millis,balents_2chain,lin98} However, very
little work \cite{lin98,konik98} concerned the close vicinity of
half--filling, a situation of considerable interest, in particular in view
of the physics of doped spin ladders. In the present paper I wish to show
that in that situation for quite general interaction strengths the Luttinger
liquid parameter \( K_{\rho } \) becomes unity (the same as for
noninteracting fermions) and that consequently one can expect a ``d--wave''
superconducting state.  This is to be contrasted with the single--chain case
where close to half--filling one finds the universal strong coupling value
\( K_{\rho }=1/2 \) and dominant antiferromagnetic
correlations.\cite{schulz_hubbard_exact,schulz_cic2d,giamarchi_rho}

Numerical work has shown strong indications of superconductivity in Hubbard
and \( t-J \) ladders, both calculating correlation functions
directly\cite{dagotto_tjladder,noack_2chain,noack_2chain_2,troyer96,kuroki96}
and by exploiting the periodicity of the ground state energy in the presence
of a magnetic flux.\cite{hayward_2chain} Nevertheless, numerical results are
very hard to interpret in the vicinity of half--filling and numerical work
has thus not helped much in understanding the doping--induced
metal--insulator transition.

I will now summarize the current understanding of the properties of two
chains of interacting fermions coupled by interchain hopping and well away
from half--filling.  In the case of weak interactions a combination of
perturbative renormalization group calculations and bosonization
\cite{fabrizio_2chain,schulz_2chain,balents_2chain} has shown the existence
of a gap in the spin excitation spectrum and that superconducting
correlations of d type dominate. More precisely, define the superconducting
order parameter as
\begin{equation}
\label{eq:delta}
\Delta _{r}=\frac{1}{\sqrt{2}}\left( a_{1\uparrow r}a_{2\downarrow
r}-a_{1\downarrow r}a_{2\uparrow r}\right) .
\end{equation}

\noindent Here \( r \) labels sites along the chains, and \( a_{isr} \) is
the fermion annihilation operator on site \( r, \) chain \( i, \) and with
spin projection \( s \). It is appropriate to call this order parameter
``d--wave'' because in Fourier space components with transverse wavevector
\( 0 \) and \( \pi \) have opposite sign. Correlations of this order
parameter decay slowly with distance:
\begin{equation}
\label{eq:supcorr}
\left\langle \Delta ^{+}_{r}\Delta _{0}\right\rangle \approx r^{-\eta _{SCd}},
\end{equation}

\noindent with \( \eta _{SCd}=1/(2K_{\rho }) \) and \( K_{\rho } \) a number
close to unity. Powerlaw decay of correlations also exists for the \( 4k_{F} \)
component of the particle density:
\begin{equation}
\label{eq:4kf}
\left\langle n_{r}n_{0}\right\rangle \approx \cos (4k_{F}r)r^{-\eta _{4k_{F}}},
\end{equation}

\noindent and one has the \emph{scaling relation} 
\begin{equation}
\label{eq:sca}
\eta _{SCd}\eta _{4k_{F}}=1
\end{equation}

\noindent (one has \( k_{F}=\pi n/2 \), where \( n \) is the average number
of fermions per site). All other correlation functions, in particular those
representing magnetic ordering, decay exponentially with distance.

Remarkably, analogous results can also be obtained for some strong--coupling
cases:\cite{schulz_2chain} first, if correlations within a single chain are
so strong that \( K_{\rho }<1/3 \), renormalization--group generated
interchain interactions dominate over the single particle hopping, and the
resulting problem can be solved, leading to the same powerlaws
(\ref{eq:supcorr},\ref{eq:4kf}) and the same scaling law (\ref{eq:sca}) as
in the weak coupling case. However, because now \( K_{\rho }<1/3 \), the \(
4k_{F} \) CDW correlations dominate over d--wave
superconductivity. Secondly, a particular limit of the two--chain \( t-J \)
model, to be discussed in more detail below, also leads to the same
powerlaws and the same scaling relation. The natural conclusion from these
findings in three different limits is that \emph{the correlated two--chain
model is in the same phase, characterized by
eqs.(\ref{eq:supcorr},\ref{eq:4kf},\ref{eq:sca}) in a large region of
interaction strength, both for weak and for strong correlation.}

I now wish to determine the value of the Luttinger liquid parameter \(
K_{\rho } \) in the vicinity of the doping--induced metal--insulator
transition, i.e. for \( n\rightarrow 1 \) but \( n\neq 1 \). The direct use
of the renormalization group approach is highly impractical because in this
case umklapp interactions play an important role at high energies, but drop
out at low energies. At intermediate energies, one then passes through a
strong--coupling region which is impossible to treat systematically. The
``\( K_{\rho }<1/3 \)'' approach is equally impossible because for \(
n\rightarrow 1 \) in a single chain one universally has \( K_{\rho
}\rightarrow 1/2 \),\cite{schulz_hubbard_exact,schulz_cic2d,giamarchi_rho}
i.e. one drops outside the validity range of this approach. However, the \(
t-J \) model gives us the possibility to reach some exact and general
conclusions.  To be specific I consider the spatially anisotropic ``\(
t-J-J_{\perp } \) model'' \cite{troyer96} with Hamiltonian
\begin{eqnarray}
H & = & -t\sum _{i,r}(a^{+}_{isr}a_{isr+1}+H.c.)-t_{\perp }\sum
 _{r}(a_{1sr}^{+}a_{2sr}+H.c.)\nonumber \\ & & +J\sum
 _{i,r}\mathbf{S}_{ir}\cdot \mathbf{S}_{ir+1}+J_{\perp }\sum
 _{r}\mathbf{S}_{1r}\cdot \mathbf{S}_{2r}\quad ,\label{eq:tjj}
\end{eqnarray}

\noindent where \textbf{\( \mathbf{S}_{ir}=a^{+}_{isr}a_{is'r}(\boldsymbol
{\sigma })_{ss'}/2 \)} is the spin operator on site \( (i,r) \) and the
usual no--double--occupancy constraint is imposed. \( J \) and \( J_{\perp }
\) are the exchange constants along and perpendicular to the ladder, and \(
t \) and \( t_{\perp } \) the corresponding hopping integrals.

\begin{figure}
{\centering \resizebox*{0.8\textwidth}{!}{\includegraphics{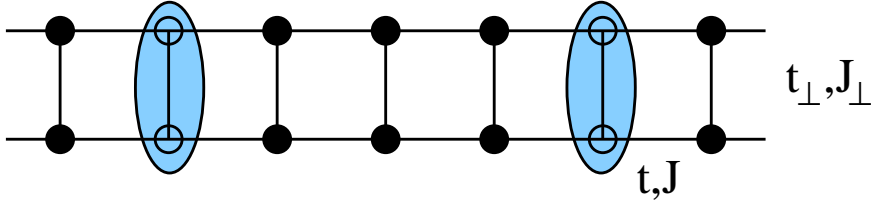}} \par}

\caption{Low--energy Hilbert space of the \protect\( t-J-J_{\perp }\protect
\) model in the limit \protect\( J_{\perp }\gg t,t_{\perp },J\protect \):
rungs are either doubly occupied and in a singlet state (full dots), or both
sites are empty (shaded ellipses).\label{fig.gs}}
\end{figure}

In the limit where \( J_{\perp }\gg J,t,t_{\perp } \) analytical progress
can be made:\cite{troyer96} then the low--energy Hilbert space consists of
rungs of the ladder where either both sites are occupied and form a singlet
or both sites are empty (fig.\ref{fig.gs}). Singly occupied sites or rungs
with a triplet lie higher by an energy of order \( J_{\perp } \). It is then
convenient to consider the state where all rungs are occupied by singlets as
the physical vacuum and creation of an empty rung as creation of a boson on
site \( r \), with associated boson creation operator \( b^{+}_{r}=\Delta
_{r} \). In second order perturbation theory in \( t \) and \( J \) one then
obtains an effective Hamiltonian for the low--energy Hilbert space:
\begin{equation}
\label{eq:heff}
H_{\mathrm{eff}}=-t_{\mathrm{eff}}\sum
_{r}(b_{r}^{+}b_{r+1}+H.c.)+V_{\mathrm{eff}}\sum _{r}n_{r}n_{r+1}\quad ,
\end{equation}

\noindent where \( t_{\mathrm{eff}}=8t^{2}/3J_{\perp } \), \(
V_{\mathrm{eff}}=(16t^{2}/3-3J^{2}/8)/J_{\perp } \), \( n_{r}=b_{r}^{+}b_{r}
\), and now there is the ``hardcore constraint'' \( n_{r}\leq 1 \). Because
of that constraint, one can straightforwardly transform the model into a
spin-\( 1/2 \) language by setting \( S_{r}^{+}=b_{r}^{+} \) and \(
S_{r}^{z}=n_{r}-1/2 \), to recover the well--known spin-\( 1/2 \) XXZ spin
chain.\cite{yang_xxz,haldane_xxzchain} In Haldane's paper the interesting
Luttinger liquid parameter \( K_{\rho } \) is determined. In particular, if
the original fermionic model is close to the insulating state at
half--filling, \( n\rightarrow 1 \), corresponding to a very dilute hardcore
boson gas, one has \( K_{\rho }=1+O(1-n) \), independent of the value of the
interaction, i.e. \( K_{\rho } \) \emph{takes a universal value when the
metal--insulator transition is approached}. Moreover, I have argued above
that both the weak--interaction limit of the two--chain model and the
strongly interacting chains weakly coupled by interchain hopping are in the
same phase as the \( t-J-J_{\perp } \) model.  One thus concludes that upon
approaching the doping induced metal--insulator transition, \( n\rightarrow
1 \), the Luttinger liquid parameter \( K_{\rho } \) takes the universal
value unity, independently on whether correlations are weak or strong. Via
the relations (\ref{eq:supcorr},\ref{eq:4kf},\ref{eq:sca}) this then implies
that \emph{d--type superconducting correlations are strongly dominant close
to the metal--insulator transition.}

It is interesting to understand how the powerlaws
(\ref{eq:supcorr},\ref{eq:4kf}) connect to the behavior expected for the
insulating case. Consider first the \( 4k_{F} \) CDW correlations: at
distances shorter than the average spacing between bosons one expects the
behavior typical of an insulator, i.e. a constant.  A reasonable form for
the crossover between short--distance insulating behavior and the
asymptotics of eq.(\ref{eq:4kf}) then is\cite{schulz_la}
\begin{equation}
\label{eq:4kf2}
\left\langle n_{r}n_{0}\right\rangle \approx \cos (4k_{F}r)(\ell
^{2}+r^{2})^{-K_{\rho }},
\end{equation}

\noindent where \( \ell \propto 1/(1-n) \) is a length proportional to the
average distance between bosons. On the other hand, the asymptotics
(\ref{eq:supcorr}) of the pairing correlations are expected to be valid only
at distances larger than \( \ell \), with exponential decay at shorter
distances. This in particular implies that the amplitude of the divergence
of the pairing susceptibility (\( \chi _{SCd}\propto \max (\omega
,q,T)^{-3/2} \) as \( K_{\rho }\rightarrow 1 \)) vanishes as \( n\rightarrow
1 \).

Recently, Konik \emph{et al.}\cite{konik98} have investigated the vicinity
of the metal--insulator transition in the weak--coupling limit, starting
from the \( SO(8) \) model of the insulating state\cite{lin98} and using the
exact integrability of that model. They arrive at the same conclusion, \(
K_{\rho }=1+O(1-n). \) This provides a confirmation of the continuity
between weak and strong correlation conjectured before.\cite{schulz_2chain}

I finally comment on some properties of the model (\ref{eq:tjj}) for general
fermion density. First, in the very dilute limit, corresponding to a dense
bosonic model (\ref{eq:heff}), from Haldane's results one again has \(
K_{\rho }=1+O(n). \) More interestingly, for a quarter--filled fermionic
model (one boson per two sites), there is a metal--insulator transition into
an insulating CDW state with period \( 4k_{F} \) when \(
V_{\mathrm{eff}}>2t_{\mathrm{eff}} \). In the XXZ spin chain, this
transition is due to umklapp processes \cite{haldane_xxzchain} which in the
original fermionic language of eq.(\ref{eq:tjj}) correspond to four--fermion
umklapp processes, as first discussed in ref.\cite{schulz_la}, and which can
become relevant only for strongly repulsive interactions. In the conducting
state (\( V_{\mathrm{eff}}<2t_{\mathrm{eff}} \)) one has \( K_{\rho }\geq
1/2 \), i.e. superconductivity dominates. On the other hand, for \(
V_{\mathrm{eff}}>2t_{\mathrm{eff}} \), when \( n\rightarrow 1/2^{\pm } \)one
has \( K_{\rho }\rightarrow 1/4, \) i.e. \( 4k_{F} \) CDW fluctuations
dominate. The \( t-J-J_{\perp } \) model thus provides an interesting
example for the transition between dominant superconducting and CDW
fluctuations. Finally, we remark that in the case of strongly attractive
interactions, \( V_{\mathrm{eff}}<-2t_{\mathrm{eff}} \), one has phase
separation for any filling of the band.

In conclusion, based on a solvable limit of the \( t-J \) ladder model and a
continuity conjecture between strong and weak correlation, I have shown here
that close to the doping--induced metal--insulator transition the Luttinger
liquid parameter \( K_{\rho } \) of the two--chain model \emph{takes the
universal value unity,} corresponding to dominant d--type superconducting
correlations.  This contrasts strongly with the single--chain case, where \(
K_{\rho }\rightarrow 1/2 \) and antiferromagnetism dominates. It should
however be pointed out that superconductivity is rather easily destroyed by
disorder,\cite{orignac_2ch_disorder} and that therefore in any real system
the existence or not of superconductivity will depend crucially on the
interplay between disorder and interladder coupling which stabilizes
superconductivity.

\end{document}